\documentclass{ar-1col-S2O}
\usepackage[numbers,sort&compress]{natbib}
\usepackage{url}
\usepackage{amsmath,amsfonts,amssymb}
\usepackage{upgreek}
\usepackage{booktabs}
\usepackage{setspace}
\usepackage{comment}

\usepackage[dvipsnames]{xcolor}
\usepackage{upgreek}

\newcommand{\cevns}{CE$\upnu$NS\,}
\newcommand{\eh}{e$^-$/h$^+$\,}
\newcommand{\epixday}{${\rm e}^-$/pixel/day}

\jname{Xxxx. Xxx. Xxx. Xxx.}
\jvol{AA}
\jyear{YYYY}
\doi{10.1146/((please add article doi))}

\begin{document}

\markboth{Baxter et al.}{Low Energy Backgrounds}

\title{Low-Energy Backgrounds in Solid-State Phonon and Charge Detectors}

\author{Daniel Baxter,$^{1,}$$^2$ Rouven Essig,$^3$ Yonit~Hochberg,$^{4,}$$^5$ Margarita Kaznacheeva,$^6$ Belina von Krosigk,$^{7,}$$^8$ Florian Reindl,$^{9,}$$^{10}$ Roger~K.~Romani,$^{11}$ and Felix Wagner$^{12,}$$^{13}$
\affil{$^1$Quantum Division, Fermi National Accelerator Laboratory, Batavia, IL 60510, USA; email: dbaxter9@fnal.gov}
\affil{$^2$Department of Physics and Astronomy, Northwestern University, Evanston, IL 60208, USA}
\affil{$^3$C.N. Yang Institute for Theoretical Physics, Stony Brook University, NY 11794, USA; email: rouven.essig@stonybrook.edu}
\affil{$^4$Racah Institute of Physics, Hebrew University of Jerusalem, Jerusalem 91904, Israel; email: yonit.hochberg@mail.huji.ac.il}
\affil{$^5$Laboratory for Elementary Particle Physics,
 Cornell University, Ithaca, NY 14853, USA}
\affil{$^6$Physics Department, TUM School of Natural Sciences, Technical University of Munich, 85747 Garching, Germany; email: margarita.kaznacheeva@tum.de}
\affil{$^7$Kirchhoff-Institute for Physics, Heidelberg University, 69120 Heidelberg, Germany; email: bkrosigk@kip.uni-heidelberg.de}
\affil{$^8$Institute for Astroparticle Physics, KIT, 76131 Karlsruhe, Germany}
\affil{$^9$HEPHY, Austrian Academy of Sciences, A-1010 Vienna, Austria}
\affil{$^{10}$ATI, TU Wien, A-1020 Vienna, Austria; email: florian.reindl@tuwien.ac.at}
\affil{$^{11}$Department of Physics, University of California, Berkeley, CA 94720, USA; email: rkromani@berkeley.edu}
\affil{$^{12}$Department of Physics, ETH Zurich, CH-8093 Zurich, Switzerland}
\affil{$^{13}$ETH Zurich - PSI Quantum Computing Hub, Paul Scherrer Institute, CH-5232 Villigen, Switzerland; email: felix.wagner@phys.ethz.ch}
}

\begin{abstract}
Solid-state phonon and charge detectors probe the scattering of weakly interacting particles, such as dark matter and neutrinos, through their low recoil thresholds. 
Recent advancements have pushed sensitivity to eV-scale energy depositions, uncovering previously-unseen low-energy excess backgrounds. 
While some arise from known processes such as thermal radiation, luminescence, and stress, others remain unexplained. 
This review examines these backgrounds, their possible origins, and parallels to low-energy effects in solids. Their understanding is essential for interpreting particle interactions at and below the eV-scale. 
\end{abstract}

\maketitle

\tableofcontents


\section{INTRODUCTION}
The rapid advancement of low-threshold detectors over the last decade has unlocked energy ranges that were previously inaccessible, which has in turn revealed new backgrounds that were not well-understood. 
The way that these effects manifest in different detectors varies, but in general, all detectors with thresholds that are sufficiently low to be sensitive to energies below $\sim$ 100\,eV measure a steeply-rising event rate with decreasing energy that we will refer to here as a low-energy excess~(LEE)~\cite{Fuss:2022fxe}. 
While there are commonalities in the various LEEs measured by different experiments using similar detector technologies, a common origin to explain all LEE rates, for example a dark matter  (DM) signal, is inconsistent with the data. 
Furthermore, the origins of the LEEs in phonon-sensitive detectors and charge-sensitive detectors are distinct. 
The modeling and mitigation of these LEEs is crucial for the continued advancement of rare event detection at the eV-scale, in particular for the direct detection of sub-GeV DM and for the precision measurement of coherent elastic neutrino-nucleus scattering (\cevns).

An LEE has been observed in various solid-state DM and \cevns detectors for at least a decade, but often at detector thresholds where systematic modeling was poor and background simulations had not yet been validated. 
The first unambiguous observations of an LEE came from the EDELWEISS~\cite{EDELWEISS:2019vjv} and CRESST~\cite{CRESST:2019jnq} Collaborations in 2019. 
In a 6-day surface run with a 33.4\,g Ge bolometer, EDELWEISS observed an LEE above a flat radiogenic background rate of $\sim 10^3$\,dru ($\equiv$\,counts/kg/day/keV) down to a 60\,eV threshold~\cite{EDELWEISS:2019vjv}. 
Shortly after, the CRESST-III experiment, operated deep underground at the Laboratori Nazionali del Gran Sasso~(LNGS), released their first low-threshold run from a 23.6\,g CaWO$_4$ crystal operated for more than a year (3.64\,kg-day total exposure after cuts)~\cite{CRESST:2019jnq}; in these low-background data, they observed a significant LEE below 200\,eV,~down to a 30\,eV analysis threshold.    
Notably, both of these cryogenic detectors measure the total energy deposited in the crystal by observing the total phonon energy following an event. 

In light of these convincing LEE observations, prior results sensitive to sub-keV energy depositions were revisited, and what had previously been widely considered as `systematic effects near threshold' were considered as newly suspect. 
This included earlier results from a CRESST-style prototype detector for \cevns~detection, NUCLEUS~\cite{CRESST:2017ues,NUCLEUS1g:2017}, which in 2017 marked the first measurement of a cryogenic LEE in the literature. 
The reinterpretation also included charge-sensitive detectors where single-electron rates had long exceeded theoretically-predicted levels, including SuperCDMS-HVeV~\cite{SuperCDMS:2018mne} and SENSEI at MINOS~\cite{SENSEI:2019ibb}; follow-up measurements from DAMIC at SNOLAB~\cite{DAMIC:2019dcn} and EDELWEISS~HV~\cite{EDELWEISS:2020fxc} displayed similar behavior. 
Around this time, attempts to find a unifying explanation for these LEEs~\cite{Kurinsky:2020dpb} were unsuccessful~\cite{Harnik:2020ugb,Knapen:2020aky} but yielded some interesting ideas on testing the nature of the LEEs~\cite{Heikinheimo:2021syx,Abbamonte:2022rfh}. 
With a single common explanation ruled out, the DM and \cevns~fields embarked on a multiyear campaign to share data and expertise to categorize and eliminate LEE backgrounds---an effort encapsulated by the EXCESS Workshop Series~\cite{Fuss:2022fxe}. 
While outside the scope of this review, gaseous and liquid detectors have long suffered from LEE backgrounds in charge-sensitive~\cite{Essig:2012yx, DarkSide:2022knj, PandaX:2022xqx, XENON:2024tjg} and heat~\cite{PICO:2016kso} channels.
There is a near-total overlap of the statistical handling of such backgrounds and a similar narrative of long-lived metastable states releasing energy over time. 

In this review, we summarize the status of the various solid-state LEEs, including what is known about their behavior in different types of detectors, what remains to be understood, and ongoing efforts towards modeling and mitigation. 
Without a comprehensive understanding, the various sources of LEEs will severely hinder the low-threshold DM and \cevns detection programs over the next decade. Beyond these two applications, there is increasing evidence that LEEs may present a challenge for solid-state quantum computing technologies, which are intrinsically sensitive to any type of environmental interaction.


\section{LOW-THRESHOLD DETECTION}

\begin{figure}[ht]
    \centering
    \includegraphics[width=1.\linewidth]{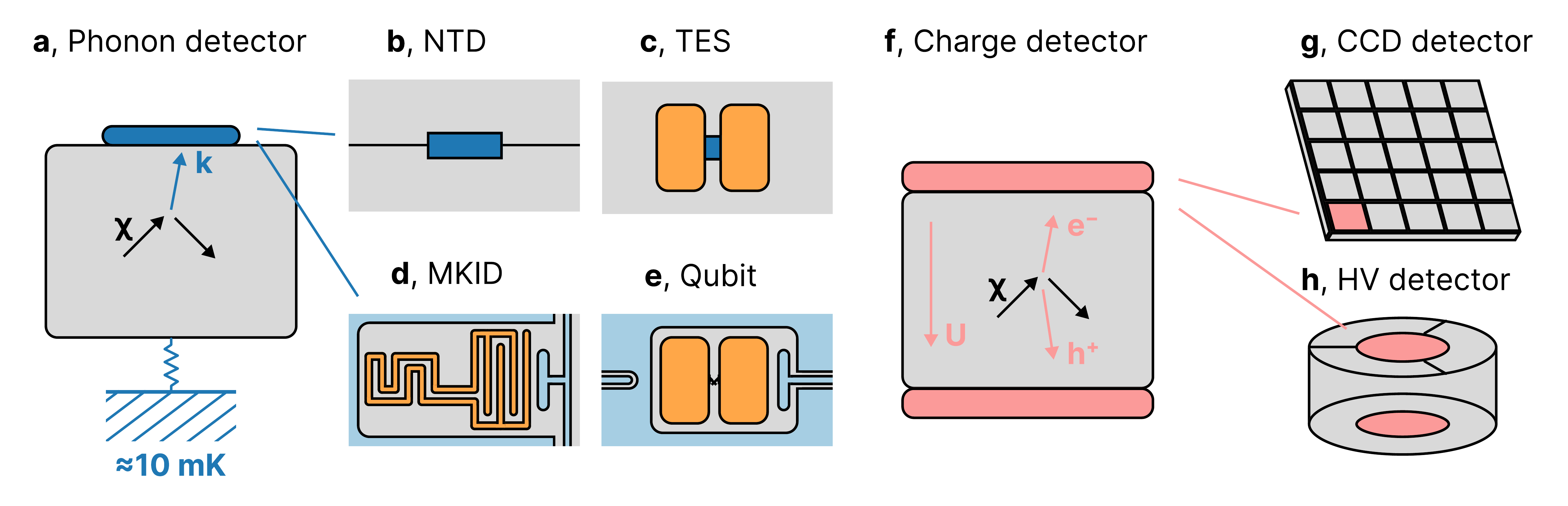}
    \caption{ {\bf Relevant detectors.} Phonon-based detectors {\bf (a)}, including NTDs {\bf (b)}, TESs {\bf (c)}, MKIDs {\bf (d)}, and qubits {\bf (e)}, wherein interactions create phonons which are registered by a phonon sensor (blue); Charge-collecting detectors {\bf (f)}, including CCD detectors {\bf (g)} and HV detectors {\bf (h)}, wherein particle interactions in the crystal bulk (gray) will generate \eh pairs that are drifted by an electric field towards electrodes on the top or bottom (red). 
    }
    \label{fig:low_threshold_detectors}
\end{figure}

The search for light (sub-GeV mass) DM particles and the study of \cevns~ kinematically requires detectors that are capable of precisely measuring eV-scale energy transfers to their bulk material. 
DM and neutrinos can transfer energy to the target nuclei or the target electrons, and (depending on the material) produce electrons, scintillation photons, and phonons~\cite{Freedman:1973yd,Essig:2011nj,Essig:2015cda,Derenzo:2016fse,Ibe:2017yqa,Knapen:2017ekk,Kurinsky:2019pgb,Griffin:2020lgd}. 
In this review, we focus on low-threshold experiments utilizing solid-state monocrystalline targets with either phonons or charge as their primary signal channel. 
Some phonon-based experiments combine their primary channel with a secondary channel (phonon plus charge/scintillation) for means of particle discrimination between electron and nuclear recoils and/or for surface-event rejection. 
For an overview on particle discrimination and surface-event rejection, we refer the reader to Refs.~\cite{kaznacheeva2024scintillatinglowtemperaturecalorimetersdirect,agnese_demonstration_2013,armengaud_performance_2017}.
Fig.~\ref{fig:low_threshold_detectors} illustrates the working principle of the relevant detectors.

\subsection{Measuring phonons}
\label{subsec:phonon}

Energy depositions of sufficient magnitude in a cryogenic crystal absorber create a population of non-thermal phonons. These decay instantaneously ($\sim$\,ns) to acoustic phonons of the Debye frequency ($\hbar \omega \sim 80$\,meV in Si) and lower, which propagate ballistically through the crystal. Within $\sim\upmu {\rm s}$ they decay to thermal energies, leading to a temperature increase of the crystal. 
The crystal then cools back to its equilibrium temperature through a thermal link with a heat bath~\cite{Proebst1995}. Sensing this phonon population is fundamentally limited by the energy of thermal phonons (thermal noise). 
Many characteristics of detectors that crucially influence their sensitivity are strongly temperature-dependent, {\it e.g.}~heat capacities and thermal couplings ($\propto T^3$ for insulators and $\propto T$ for metals), and the superconducting state and its equilibrium quasiparticle (QP) population. Phonon sensors are therefore typically operated at temperatures of about 10\,mK in dilution refrigerators. 
Typical heat capacities of the system are on the order of a few pJ/mK, and thermal couplings are on the order of several hundreds of pW/mK, leading to pulse lengths of ms to hundreds of ms.  
The initially deposited energy can originate from the recoil of a massive particle or radiation interactions in the crystal, sensor, or their interface, but also from stress or vibrations. 
Sensing this energy can follow two strategies depicted in Fig.~\ref{fig:low_threshold_detectors}a: either collecting the prompt athermal phonon population or measuring the temperature increase. We discuss several sensor types in what follows.

A Neutron Transmutation Doped (NTD) sensor (Fig.~\ref{fig:low_threshold_detectors}b) is a semiconductor thermistor, often using Ge, which is strongly coupled to the absorber crystal, sensing the thermal phonon population. 
Its sensitivity and timing resolution depend strongly on the heat capacity and thermal coupling of the absorber crystal. Typical energy thresholds of such sensors are $>$\,keV and timing resolutions on the order of 100\,ms to seconds \cite{armengaud_performance_2017}. 
Transition Edge Sensors (TESs) are superconducting films operated in transition to their normal conducting state (Fig.~\ref{fig:low_threshold_detectors}c). The strong temperature dependence of the electrical resistance in this state enables sensitivity to $\sim \upmu {\rm K}$ temperature fluctuations.
Additional fully-superconducting phonon collectors can amplify the signal: the athermal phonon population generates Bogoliubov QPs in the collector that are successively trapped in the TES, improving the phonon collection efficiency without adding to the heat capacity or thermal coupling. 
Optimized sensors can reach energy thresholds of $\sim$\,eV and sub-ms timing resolutions \cite{CRESST:2024cpr, SuperCDMS:2020aus}. 
The combination of tungsten TES (W-TES) and aluminum fins used in such detectors is known as a Quasiparticle-trap-assisted Electrothermal-feedback TES (QET)~\cite{Irwin1995}.

The creation of QPs in superconductors can be leveraged as a fully athermal readout scheme. 
This principle is used in Microwave Kinetic Inductance Detectors (MKIDs), superconducting resonators with frequencies in the GHz-regime (Fig.~\ref{fig:low_threshold_detectors}d). 
The presence of QPs shifts the resonator frequency, resulting in a measurable signal. 
MKIDs have achieved energy thresholds $< 100$\,eV and timing resolutions of about 100\,$\upmu$s \cite{Delicato:2023wrg,Temples:2024ntv}. Alternatively, a Superconducting Quantum Interference Device (SQUID)-capacitor circuit can be operated as a qubit (Fig.~\ref{fig:low_threshold_detectors}e), keeping a coherent quantum state for $\gtrapprox 100\, \upmu {\rm s}$ that is read out and manipulated through microwave pulses. 
The coherence and transition energies of such qubits are sensitive to single QPs tunneling through the Josephson junctions, which has been studied extensively in the quantum computing literature \cite{PhysRevB.84.064517, PRXQuantum.3.040304, PhysRevB.110.024519}. Prototypes of using such qubits as sensitive particle detectors have been studied, with optimized designs proposed \cite{PhysRevApplied.22.054009, Linehan:2024niv, ramanathan2024quantumparitydetectorsqubit}. 

While the calibration of low-threshold phonon detectors is challenging, multiple methods have been employed. Calibrations using both electronic energy deposition ($\sim$ 6 keV $\gamma$ rays from $^{55}$Fe, e.g. \cite{CRESST_LEE_2023, 10.1063/5.0032372}, $\sim$ eV optical photon sources, e.g. \cite{TESSERACTTwoChannel, CRESST:2024cpr}), nuclear events \cite{HVeVCalibration, CRABCalibration} and drifted charges within the detector \cite{SuperCDMS:2024yiv} have all been used to calibrate low threshold phonon detectors, with broadly self consistent results. Significantly above the trigger threshold and below the saturation point of a given detector, this calibration essentially serves to measure the linear scale factor (in some contexts, the phonon collection efficiency) between a given energy deposited in the detector's phonon system and the size of the observed signal. Broadly, the response of these low-threshold phonon detectors seems to be well understood.

\subsection{Measuring charges}
\label{subsec:charge_production}

Traditional charge-sensitive solid-state DM detectors are made from either Si or Ge. 
When energy is deposited in the semiconductor crystal, it can excite electrons from the valence band to the conduction band, creating electron/hole (\eh) pairs. 
The energy required to create one \eh pair in a semiconductor is typically larger than the bandgap because some energy will be lost to phonons. 
The average energy required to create an \eh pair at room temperature is $\sim 3.6$\,eV for Si and $\sim 2.9$\,eV for Ge.
Using an applied electric field, \eh pairs in the bulk are drifted to the edges of the detector, where they are measured, as shown in Fig.~\ref{fig:low_threshold_detectors}f. 
The charge collection efficiency and number of measured charge carriers are influenced by  factors such as impurities, defects, and drift field strength.
If the electric field is weak or the time for the carriers to travel across the crystal is long, an electron from the conduction band can recombine with a hole in the valence band, resulting in the loss of both charge carriers and reducing the detectable signal. 
The ionization yield is higher for electron recoils compared to nuclear recoils. 
We therefore discuss the energy measured by a charge detector in units of electron-equivalent energy, (eV$_{\rm ee}$), as the recoil energy valid for an electron interaction, to distinguish from phonon detectors which measure the total energy deposited independently of the type of interaction.

A Charge-Coupled Device (CCD) consists of many coupled capacitors, effectively dividing up the detector into pixels, as shown in Fig.~\ref{fig:low_threshold_detectors}g.  Any \eh pairs produced in the bulk are drifted towards a `buried channel', an extremum of the potential close to the detector surface.  
This channel can be p-type, in which case it stores holes, or n-type, in which case it stores electrons. 
Gates on either the front- or backside of the device move the charges in each capacitor to its neighbor through synchronized clocked voltages, until they reach the `serial register', a row of pixels across which the charges can be transferred to the readout amplifier.  
In a conventional CCD, the total charge in each pixel is measured once before it is discarded. 
Electronic  noise from the CCD output amplifier adds both high- and low-frequency readout noise. 
High-frequency readout noise can be reduced by correlated double sampling~\cite{janesick2001scientific}, but low-frequency noise remains and results in a measurement uncertainty for each pixel charge of $\sim$ $2-3$~\eh rms/pixel.  
In a Skipper-CCD, the readout amplifier is modified by coupling a floating gate output stage to a small-capacitance sense-node.  
In this way, multiple non-destructive measurements of the same pixel charge can be performed, reducing also the low-frequency noise and allowing a measurement of pixel charge to a precision of $\ll 1$\,\eh\,rms/pixel.

In the case of cryogenic charge-sensitive devices as depicted in Fig.~\ref{fig:low_threshold_detectors}h, two scenarios should be distinguished.
In the first, the applied bias voltage is on the  order of a few V, resulting in an electric field strength of $\lesssim 1$\,V/cm.
Electrodes on the crystal surfaces, instrumented with high-impedance charge amplifiers shown in Fig.~\ref{fig:low_threshold_detectors}h, measure the image charge collected, providing a direct measure of the deposited energy used to produce \eh pairs.
Alternatively, larger voltages on the order of 100\,V can be applied, resulting in electric field strengths of $> 100$\,V/cm; such devices are commonly referred to as high-voltage (HV) detectors.
The purpose of the HV is to amplify the charge signal in the form of secondary phonons, exploiting the Neganov–Trofimov–Luke (NTL) effect \cite{Neganov:1985khw,Luke:1988xcw}. 
As \eh pairs drift across the crystal, they quickly reach terminal velocity, long before reaching a surface.
Due to significant charge-phonon interactions, the excess work done on the carriers is transferred to the crystal lattice in the form of NTL phonons.
Consequently, the measured total phonon energy utilizing sensors described in Sec.~\ref{subsec:phonon} scales directly with the bias voltage and provides a strongly-amplified phonon signal that dwarfs the original charge signal.
With bias voltages $\sim 100$\,V, the total NTL phonon energy per initial \eh pair is $\sim 100$\,eV, allowing for resolutions $\ll$~1\,\eh and functionally enabling quantized \eh pair readout.


\section{SUMMARY OF OBSERVED LOW-ENERGY EVENTS}

\begin{figure}[ht]
    \centering
    \includegraphics[width=1.\linewidth]{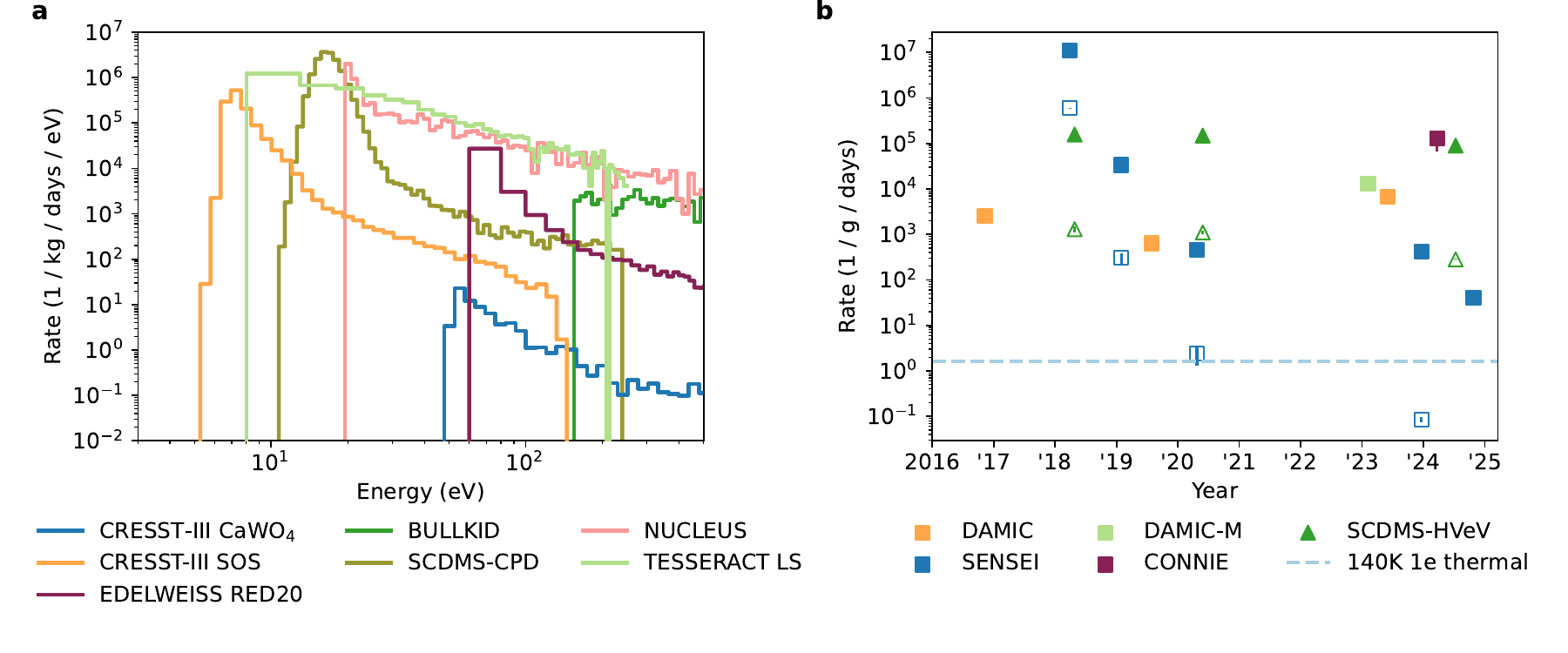}
    \caption{ {\bf Low-energy excess rates.} {\bf (a)} Spectra observed by various phonon-based cryogenic experiments: CRESST-III (CaWO$_4$~\cite{CRESST_LEE_2023} and Si-On-Sapphire (SOS)~\cite{CRESST:2024cpr}, W-TES,  LNGS), EDELWEISS RED20 (Ge, NTD, surface)~\cite{EDELWEISS:2019vjv}, BULLKID (Si, MKID, surface)~\cite{BULLKID2022}, SuperCDMS-CPD (Si, W-TES, surface)~\cite{SuperCDMS:2020aus}, NUCLEUS (Al$_2$O$_3$, W-TES, surface)~\cite{CRESST:2017ues}, and TESSERACT Low Stress (LS) configuration (Si, W-TES, surface)~\cite{anthonypetersen2022stressinducedsourcephonon}. 
    {\bf (b)} Single (double) electron rates in Si shown with solid (open) markers for each experiment with sensitivity to individual \eh production; data from DAMIC~\cite{DAMIC:2016qck,DAMIC:2019dcn,SENSEI:2023rcc}, SENSEI~\cite{Crisler:2018gci,SENSEI:2019ibb,SENSEI:2020dpa,SENSEI:2023zdf,SENSEI:2024yyt}, DAMIC-M~\cite{DAMIC-M:2023gxo}, CONNIE~\cite{CONNIE:2024pwt}, and SuperCDMS-HVeV~\cite{SuperCDMS:2018mne,SuperCDMS:2020ymb,SuperCDMS:2024yiv} are organized by the date of first arXiv posting; square (triangle) points indicate CCD (cryogenic) experiments at $\sim$140\,K ($<1$\,K), and the dashed line indicates an upper limit of dark rate from bulk thermal single-electron excitations for a CCD operated at 140\,K according to Eq.~\ref{eq:thermal}.
} 
    \label{fig:excess_rates}
\end{figure}

\noindent LEE backgrounds have been observed by multiple low-threshold experiments. 
These experiments are either optimized to sense charge or phonon signals, with some overlap of experiments that are sensitive to both. 
In the following, we discuss the LEEs observed in phonon detectors in Sec.~\ref{ssec:phonon_excesses} and those in charge detectors in Sec.~\ref{ssec:charge_excesses}.

\subsection{Low-energy events in phonon detectors}
\label{ssec:phonon_excesses}

\definecolor{softgreen}{HTML}{96bf9a} 
\definecolor{softred}{HTML}{ffa387}   
\begin{table}[t]
\caption{Common properties of LEE background observed in various experiments. 
    }
    \centering
    \begin{tabular}{|c||c|c|c|c|c|}\hline
       \textbf{\textcolor[HTML]{0c87cc}
       {Experiment}} & \textbf{\textcolor[HTML]{0c87cc}{(1) rise}} & \textbf{\textcolor[HTML]{0c87cc}{(2) decay}} & \textbf{\textcolor[HTML]{0c87cc}{(3) ion.}} & \textbf{\textcolor[HTML]{0c87cc}{(4) coin.}} & \textbf{\textcolor[HTML]{0c87cc}{(5) absorber}} \\
       \hline \hline
       CRESST-III \cite{CRESST_LEE_2023,CRESST:2TES2024}& \textcolor{softgreen}{yes} & \textcolor{softgreen}{yes} & N.A. & \textcolor{softred}{no} & \textcolor{softgreen}{yes} \\
       NUCLEUS \cite{NUCLEUS1g:2017, 2tes_nucleus_workshop24}& \textcolor{softgreen}{yes} & \textcolor{softgreen}{yes} & N.A. & N.A. & \textcolor{softgreen}{yes} \\
       TESSERACT \cite{anthonypetersen2022stressinducedsourcephonon,TESSERACTTwoChannel}& \textcolor{softgreen}{yes} & \textcolor{softgreen}{yes} & N.A. & \textcolor{softred}{no} & \textcolor{softgreen}{yes} \\
       SuperCDMS-CPD \cite{SuperCDMS:2020aus, phdthesis_Underwood_2022}& \textcolor{softgreen}{yes} & \textcolor{softgreen}{yes} & N.A. & N.A. & N.A. \\
       EDELWEISS \cite{EDELWEISS:2020fxc, queguiner_edelweiss_thesis_2018,EDELWEISS:2023hcg}& \textcolor{softgreen}{yes} & \textcolor{softgreen}{yes} & \textcolor{softred}{no} & N.A. & \textcolor{softgreen}{yes} \\
       RICOCHET \cite{Ricochet2023}& \textcolor{softgreen}{yes} & N.A. & \textcolor{softred}{no} & N.A. & N.A. \\
       BULLKID \cite{Delicato:2023wrg}& \textcolor{softred}{no} & N.A. & N.A. & N.A. & N.A. \\
       Mannila et al. \cite{Mannila2022} & \textcolor{softgreen}{yes} & \textcolor{softgreen}{yes} & N.A. & N.A. & N.A. \\
       \hline
    \end{tabular}
    
    \label{tab:phonon_excesses}
\end{table}

Phonon LEE backgrounds were measured in different sensor types, with various materials, and at setups with strongly-varying background levels. These measurements exhibit several properties which are commonly observed in various experiments, as summarized in Tab.~\ref{tab:phonon_excesses}, and which are detailed in the following:

\begin{itemize}
\item[(1)] The excess background rate rises steeply towards lower energies, typically observable below $\sim100-200$\,eV. In most data sets, the rise of the background does not follow a single exponential or power law model, but requires combinations thereof.

\item[(2)] The excess background rate monotonically decays with time after the cooldown of the setup, with subsequent warm up cycles partially or completely resetting the rate.

\item[(3)] Detectors sensitive to both phonons and charges observe excess background events only in the form of phonons.
The background source(s) thus appear to couple exclusively to the phonon system, with no simultaneous ionization.

\item[(4)] Detectors placed in a joint housing, {\it i.e.}~without light-tight shielding against each other, do not observe the excess background events in coincidence.

\item[(5)] Detectors that can discriminate between absorber and sensor events observe a sig\-ni\-fi\-cant share of the background excess events depositing energy in the phonon system of the absorber. And some events have been observed to deposit energy directly in the sensor.
 
\end{itemize}

\noindent Several recent data sets with clearly visible LEE background are shown in Fig.~\ref{fig:excess_rates}a, from the BULLKID, CRESST, NUCLEUS, TESSERACT and SuperCDMS-CPD experiments. 
We detail the measurements and experiments below.

\subsubsection{CRESST} 

The CRESST-III experiment searches for nuclear recoils induced by low-mass DM particles in absorber crystals operated at mK temperatures in a low-background setup located deep underground at the LNGS laboratories in Italy.
Particle interactions in the crystal generate phonons, which are measured by W-TESs with aluminum phonon collectors, both evaporated directly on the crystals' surface with a SiO$_2$ layer in between. 
In its first dedicated low-threshold data-taking campaign~\cite{CRESST:2019jnq}, CRESST-III operated CaWO$_4$ absorber crystals with individual masses of 24\,g. The best-performing detector achieved a nuclear recoil threshold of 30\,eV. An LEE was clearly visible in this detector, 
starting $\lesssim$\,200\,eV and rising towards the threshold. 
This observation was confirmed by four other detectors in the same run with thresholds also below 100\,eV.
Despite the presence of this LEE, leading limits were set for DM masses $< 1$\,GeV. 
For a detector of the same performance, but without an LEE, these limits would be up to two orders of magnitude stronger~\cite{CRESSTtbp}.

Consequently, studying this LEE was a top priority for a measurement campaign that ran from 2020 to 2024. Detectors with different target materials, including Al$_2$O$_3$, LiAlO$_2$ and Si, in addition to CaWO$_4$, were operated. Several modules had further modifications to their design, such as different holding structures and surrounding materials, in order to test their impact on the LEE~\cite{CRESST_LEE_2023}.
The CRESST energy spectra in Fig.~\ref{fig:excess_rates}a show that the LEE was observed across different targets, and cannot be attributed to an effect of a single material. It is observed in detectors with different geometries (wafer $20\times20\times0.4$ mm$^3$ vs.~bulky $20\times20\times10$ mm$^3$ detectors), and the count rate of the currently dominating LEE component does not seem to scale with the target mass.
Moreover, CaWO$_4$ crystals of different origins were used with 
no significant difference in the LEE rate observed~\cite{CRESST_LEE_2023}. 
In all detectors, the LEE rate decays with time since cooldown, with decay times of $\mathcal{O}(100\,\mathrm{days})$; details are given in Sec.~\ref{sec:phononorigins}. 
Triggered by the observations of EDELWEISS~\cite{Fuss:2022fxe}, CRESST performed a series of warm-up tests to different temperatures (60\,K, 600\,mK and 200\,mK); 
results are shown in Sec.~\ref{sec:phononorigins}.
CRESST finds that warm-ups to temperatures of $\mathcal{O}(10~\mathrm{K})$ or higher enhance the LEE rate, which then decays with a fast $\mathcal{O}(10\,\mathrm{days})$ component before the previously observed slow decay component of $\mathcal{O}(100\,\mathrm{days})$ resumes.
CRESST currently operates detectors with two TESs on the target crystal (`doubleTES') to disentangle events originating from a single TES or its interface to the crystal from events in the bulk \cite{CRESST:2TES2024}. Details and first prototype data are discussed in Sec. \ref{subsubsec:sensorstress}.

\subsubsection{NUCLEUS}

The NUCLEUS experiment aims to measure \cevns~from reactor anti-neutrinos~\cite{NUCLEUS:2017}.
It employs gram-scale CaWO$_4$ and Al$_2$O$_3$ cryogenic detectors, equipped with W-TES sensors to achieve low-energy thresholds. 
The LEE spectrum from a prototype measurement using a 0.5\,g Al$_2$O$_3$ detector with 20\,eV threshold operated at the surface is shown in Fig.~\ref{fig:excess_rates}a. 
The same Al$_2$O$_3$ target crystal was later operated held by two thin Si wafers, each equipped with its own TES~\cite{NUCLEUS1g:rothe2020}. This instrumented inner-veto system acts as both a surface veto and an instrumented holder. Given the expected moderate phonon transmission across the mechanical contact between the target and the inner veto, events originating in the veto wafers or on the crystal-holder interfaces are expected to exhibit different energy sharing between the target and veto TESs compared to events originating within the crystal volume, such as particle recoils. The low-energy spectrum was efficiently cleaned up using this veto cut~\cite{NUCLEUS1g:rothe2020}. 
More recently, NUCLEUS operated an Al$_2$O$_3$ target crystal directly instrumented with two TESs~\cite{2tes_nucleus_workshop24}. Similar to observations in CRESST, two distinct populations of LEE events were identified: one exhibiting equal energy sharing between the two sensors, and the other in anti-coincidence between the sensors. 

\subsubsection{SuperCDMS-CPD} SuperCDMS-CPD deployed a 10.6\,g Si target with 45.6\,cm$^2$ surface area and 1\,mm thickness, equipped with a single channel of 1031 QETs. The setup was shielded by 5\,cm copper in an above ground laboratory. An energy resolution of 3.86\,eV was reached, allowing for an analysis threshold of 16.2\,eV \cite{SuperCDMS:2020aus}. The recoil energy spectrum, shown in Fig.~\ref{fig:excess_rates}a, clearly shows a strong rise towards lower energies. 

\subsubsection{TESSERACT}

TESSERACT, also known as SPICE and HeRALD~\cite{PhysRevD.110.072006}, is a multi-target light DM experiment using QETs attached to a crystal substrate, which form either DM detectors ({\it e.g.} Al$_2$O$_3$, Si), or readout for separate DM targets ({\it e.g.} He, GaAs). 
The collaboration has demonstrated phonon energy resolutions as low as 375\,meV  in their phonon detectors, 
which measure bursts using two-channel arrays of many QETs to suppress LEEs through multi-channel coincidence~\cite{TESSERACTTwoChannel}. 
TESSERACT has shown that glued-down detectors host large LEE populations, whereas suspended detectors (\textit{e.g.} from wire bonds -- `low stress' holding) exhibit a lower residual LEE background \cite{anthonypetersen2022stressinducedsourcephonon}. The spectrum measured in the low stress configuration is shown in Fig.~\ref{fig:excess_rates}a and further discussed is Sec.~\ref{subsubsec:holdingstress}. In separate studies, the collaboration showed that this residual population was composed of two components: energy depositions in the absorber crystal (`shared' LEE) as well as energy depositions in the metal films from which the phonon sensor are constructed (`singles' LEE) \cite{TESSERACTTwoChannel}. Finally, the collaboration has seen signatures of LEE-related processes (phonon coupling, multi-channel correlation, relaxation with time) in excess noise that limits the resolution of the phonon detectors, indicating that understanding and reducing LEE backgrounds may be key to improving low-temperature sensor performance for {\it e.g.} astronomical applications and axion searches as well.

\subsubsection{EDELWEISS}
\label{sssec:edelweiss}
EDELWEISS deployed a 33.4\,g ultra-pure Ge absorber with a diameter and height of 20\,mm, sensed with a Ge-NTD thermistor of 2\,$\times$\,2\,$\times$\,0.5\,mm$^3$, glued to the top surface of the absorber.
Electrical contacts were made using Au wires bonded to the Ge-NTD on one side and to Au pads deposited on a Kapton tape glued to the copper housing of the detector on the other. 
The low-energy spectrum from this detector is shown in Fig.~\ref{fig:excess_rates}a.
By additionally reading out ionization signal, EDELWEISS demonstrated that a significant fraction of their background down to at least 15\,eV$_{\rm ee}$ comes from events that lack an ionization signal and consist of heat-only (HO) \cite{EDELWEISS:2020fxc}.
Furthermore, they were the first to observe a strong decrease of the respective excess rate after the beginning of a cooldown, discussed in Sec.~\ref{sec:phononorigins}, before reaching a steady rate.
EDELWEISS also ruled out the possibility that a single source could explain the observed HO event rate, partly due to the steady HO excess rate observed in all runs and the abrupt variations in the HO rate observed in only some of the runs.~\cite{queguiner_edelweiss_thesis_2018}.
While not covered in detail here, EDELWEISS separately operated Ge-HV charge detectors with similar results~\cite{EDELWEISS:2020fxc} to those discussed in Sec.~\ref{ssec:charge_excesses}.

\subsubsection{RICOCHET} 

RICOCHET is a planned experiment to search for \cevns~interactions in Ge detectors operated near a nuclear reactor in France. 
In the CryoCube detector concept, phonons are read out with NTD sensors glued to the detector surface, while charge is separately read out using high-electron-mobility transistor (HEMT) amplifiers. A prototype measurement operated three 38\,g Ge crystals (30\,mm diameter, 10\,mm height) at 15\,mK, achieving a 30\,eV$_{\rm ee}$ ionization energy resolution \cite{Ricochet2023}. Each crystal featured planar aluminum electrodes covering the top and bottom surfaces, with a 2×2\,mm$^2$ NTD sensor centered on the top electrode. LEE events were observed up to keV energies, significantly higher than in other experiments, and dominated by HO events as in previous EDELWEISS measurements. Using dual phonon and charge readout, RICOCHET strongly constrained ionization associated with LEE events, an observation which is now believed to hold generally for phonon LEE backgrounds.

\subsubsection{BULLKID}

BULLKID operates an array of 60 MKIDs attached to a joint 3~inch Si wafer of 5\,mm thickness. The wafer has notches in the shape of a rectangular grid, such that each sensor is most sensitive in an individual dice of 0.34\,g. A low-energy background spectrum recorded in 39\,h of surface data with an energy threshold of $160 \pm 13$\,eV is shown in Fig.~\ref{fig:excess_rates}a~\cite{Delicato:2023wrg}. After data quality and coincidence cuts between dies, a flat background of \text{$2.0 \pm 0.1 {\rm \,(stat.)} \pm 0.2 {\rm \,(syst.)} \times 10^6$\,dru} remains, with no observable rise towards lower energies. Further study will determine whether this is due to an intrinsic property of the BULLKID detector or MKIDs that makes them resilient against the sources of LEE backgrounds seen by other detectors described here, or if such LEE backgrounds are simply subdominant to the large flat background rates in the BULLKID surface detector. 

\subsubsection{Superconducting quantum devices}

QP populations exceeding the thermal state of the superconductor have also been found in superconducting quantum devices. 
QPs are created by energy depositions exceeding the binding energy of Cooper pairs, {\it e.g.}~$2 \Delta \sim 87\, {\rm GHz} \times \hbar \sim 0.36 $\,meV in aluminum films \cite{PhysRev.135.A19}. 
The thermal production of QPs at typical operation temperatures of about $10-40$\,mK is strongly suppressed, such that
one would expect to observe less than a single QP in an Earth-size aluminum volume. 
However, typical QP densities observed experimentally in aluminum are on the order of $0.01-10$/$\upmu {\rm m}^3$, as calculated from typical tunneling rates in Josephson junctions of $\sim {\rm Hz}-{\rm kHz}$. The tunneling rate depends approximately linearly on the QP density in the junction, and on the properties of the junction itself, such as dimensions and purity \cite{Kurter2022}.
The presence of such excess QPs is evidence for athermal interactions. Ref.~\cite{Mannila2022} provides a detailed study of the QP density on a $\sim \upmu$m-sized aluminum island, connected through a tunnel junction to a copper ground plane. The QPs in this study were sensed through a capacitively-coupled single-electron transistor. The measured QP density features a decaying time dependency since the cooldown of the setup, similar to the decaying LEE backgrounds measured in other low-threshold detectors. Potential origins and mitigation strategies for these excess QPs are discussed in Sec.~\ref{sec:quasiparticles}.

\subsection{Low-energy events in charge detectors}
\label{ssec:charge_excesses}

Colloquially, the measured single-electron rate in a charge-sensitive detector is often referred to as a dark rate. 
However, this rate inherently includes both the sum of true dark rate \emph{and} any LEE contributions. 
Many sources of single-electron events can come from imperfections in detector design, and not from external energy deposition or internal energy release. 
Dark rate strictly refers to single \eh pair production in the absence of a localized energy source. 
We define single-electron LEE rates as any additional sources of events on top of dark rate contributions. 
The progression of single-electron rates measured in various silicon charge detectors over time is shown in Fig.~\ref{fig:excess_rates}b.

In contrast to phonon LEEs, where possibly a single effect might have dominated LEE observations, the low-threshold charge detector community has made considerable progress in mitigating single-electron rates by identifying the following distinct contributions:
\begin{itemize}
    \item Infrared (IR) radiation, as from warm components in the same volume as the detector (such as a room temperature vacuum vessel), are able to generate single \eh pairs and leak into detector housings otherwise thought to be `light-tight'.
    \item High-energy particles from cosmic rays or background radioactivity can, through various processes, create low-energy optical photon bursts~\cite{Du:2020ldo}.
    \item CCD readout can induce spurious charge events, which can largely be mitigated through detector design and operation.
    \item Charge can thermally ionize at the relatively-high temperatures ($\sim$ 140\,K) at which CCDs are operated; this component is negligible for cryogenic HV detectors.
\end{itemize}

\subsubsection{SuperCDMS-HVeV}
\label{sssec:hvev}

The SuperCDMS R\&D HVeV detectors are sensitive to single \eh pairs by exploiting NTL amplification (see Sec.~\ref{subsec:charge_production}). 
Made of high-purity Si crystals (10\,$\times$\,10\,$\times$\,4\,mm$^3$, $0.93 \, \text{g}$), one crystal face is patterned with QETs in two readout channels to collect phonons, while the opposite face has an aluminum grid. 
The QETs are grounded, and the grid can be biased with HV ($\sim 100 \, \text{V}$), creating a uniform electric field. 
The first HVeV detector achieved a threshold of $\leq 0.8$ \eh pairs in the Run\,1 DM search~\cite{SuperCDMS:2018mne}, outperformed by the following HVeV generations.
The second generation of HVeV detectors increased the QET coverage from $13\%$ up to $50\%$ using different QET configuration masks and omitted a 40\,nm amorphous Si layer. 
Initially, it was believed that this layer would reduce the probability of charge leakage from the bias electrodes into the crystal. However, tests showed that leakage did not significantly increase when the QETs were printed directly onto the crystal surface, leading to the omission of the layer in later generations.
Both Runs 1 and 2 were conducted on the surface, at different locations and with different shielding, with the crystal clamped between two Printed Circuit Boards (PCBs) for electrical connections and heat sinking. The observed event rates were comparable in both runs~\cite{SuperCDMS:2020ymb}.

The Run\,2 detector was also used for a zero-Volt (0VeV) run to probe nuclear recoils, instead of electronic interactions, down to $\sim 10$\,eV \cite{SuperCDMS:2022zmd}.
Additional analysis of the 0VeV data led to the understanding that a contribution to the LEE is consistent with luminescence events from the PCBs (see Sec.~\ref{ssec:radiation}). 
In HVeV Run\,3, four detectors were mounted in pairs on two copper holders, operated at the Northwestern Experimental Underground Site (NEXUS) with 255\,m.w.e. overburden~\cite{SuperCDMS:2024yiv}.
Each detector was clamped between two PCBs facing the neighboring detectors.
Most of the events observed in Run\,3 were coincident in multiple detectors, suggesting that external sources are dominating over internal sources (such as mechanical stress) in this run.
Data recently taken at NEXUS with a new detector housing without PCB confirms that this background from luminescence can be successfully mitigated, reducing the LEE at higher energies.
However, these data also indicate that a sensitivity-limiting LEE is still present at the lowest energies of interest.

\subsubsection{DAMIC at SNOLAB}

The progression of CCD experiments over the years has largely focused on the reduction of single-electron rates alongside improved shielding to reduce radiogenic backgrounds. 
This began with the DArk Matter In CCDs (DAMIC) experiment, which released its first results in 2011~\cite{DAMIC:2011khz}. 
Following this result, the DAMIC Collaboration designed a larger, low-background detector named DAMIC at SNOLAB, which released its first results in 2016~\cite{DAMIC:2016lrs}. 
This experiment was then upgraded with lower-background materials and more CCDs (for a total Si mass of 42\,g), which released results for 11\,kg-days of exposure in 2020~\cite{DAMIC:2020cut}. 
Among other observations, this dataset included the first observation of a spatially-uniform bulk LEE above modeled backgrounds~\cite{DAMIC:2021crr} at energies below 200\,eV$_{\rm ee}$.  
To further investigate this LEE, the DAMIC Collaboration (working with SENSEI and DAMIC-M) upgraded DAMIC at SNOLAB with Skipper CCDs to push to lower thresholds to probe this LEE signal. 
In 2023, they confirmed the presence of the bulk LEE in a different set of CCDs with a 23~eV$_{\rm ee}$ threshold in 3.25 kg-days at $>3\sigma$ significance~\cite{SENSEI:2023rcc}. 
This LEE seems systematically different from the other LEE backgrounds in charge detectors, and a DM-type origin cannot be excluded from its own properties.

\subsubsection{SENSEI}

The Sub-Electron Noise Skipper-CCD Experimental Instrument (SENSEI) Collaboration built on the success of DAMIC by pioneering the \text{Skipper-CCD technology}, with which the charge of each pixel can be measured to sub-electron precision~\cite{Tiffenberg:2017aac}. Over several years, SENSEI has reduced the single-electron background rate from 1.14\,\epixday~\cite{Crisler:2018gci} in 2018 to $1.39\times 10^{-5}$\,\epixday~\cite{SENSEI:2024yyt} in 2024.  
This reduction, and the corresponding reduction of two-electron backgrounds, is shown in Fig.~\ref{fig:excess_rates}b and converted into mass-dependent rates to remove the effects of differing CCD thicknesses over the years. 
Gradual improvement was achieved through better shielding and reduction of light-leaks in the CCD housings.  The improved shielding reduced the environmental radiation, which was shown to produce low-energy photons from Cherenkov radiation that can create single-electron events~\cite{Du:2020ldo}.  
The most recent measurements~\cite{SENSEI:2024yyt} suggest that there is only a mild excess of single-electron events over thermal excitation backgrounds remaining at SENSEI, which is expected to be reduced further in future measurements.

In Ref.~\cite{SENSEI:2023zdf}, SENSEI observed four three-electron events, while the expectation from the pile-up of single-electron events was zero.  Two of the three-electron events appear in pixels that are likely impacted by detector defects, but the origin of the other two is unknown. More data is needed to determine their origin.
\subsubsection{DAMIC-M}
DAMIC at Modane (DAMIC-M) represents the next stage in the CCD DM experiment progression, with plans to scale up to 1\,kg of Skipper CCD mass in a 0.1\,dru low-background environment. 
Meanwhile, the DAMIC-M prototype low-background chamber (DAMIC-M LBC~\cite{DAMIC-M:2024ooa}) has already begun producing results~\cite{DAMIC-M:2023gxo}. 
Soon, the DAMIC-M LBC will accumulate sufficient exposure to test the bulk LEE observed at DAMIC at SNOLAB in a different environment, but with similar CCDs, providing a strong test of possible radiogenic origins, such as neutrons. 
Meanwhile, the LBC has achieved initial single-electron rates of $4.5 \times 10^{-3}$\,\epixday~\cite{DAMIC-M:2023hgj} with improvements expected in the next year.

\subsubsection{CONNIE}
The Coherent Neutrino-Nucleus Interaction Experiment (CONNIE) uses Skipper-CCDs for reactor \cevns detection, and has recently characterized a background (reactor-off) single-electron rate of $4.5 \times 10^{-2}$\,\epixday~\cite{CONNIE:2024pwt}. While slightly higher than the single-electron rates measured by the CCD DM experiments, this rate is notable due to the higher radiation environment present in a reactor, and provides valuable input on the radiation-induced component of single-electron LEEs.


\section{ORIGINS OF PHONON EVENTS} \label{sec:phononorigins}

\begin{figure}[ht]
    \centering
    \includegraphics[width=\textwidth]{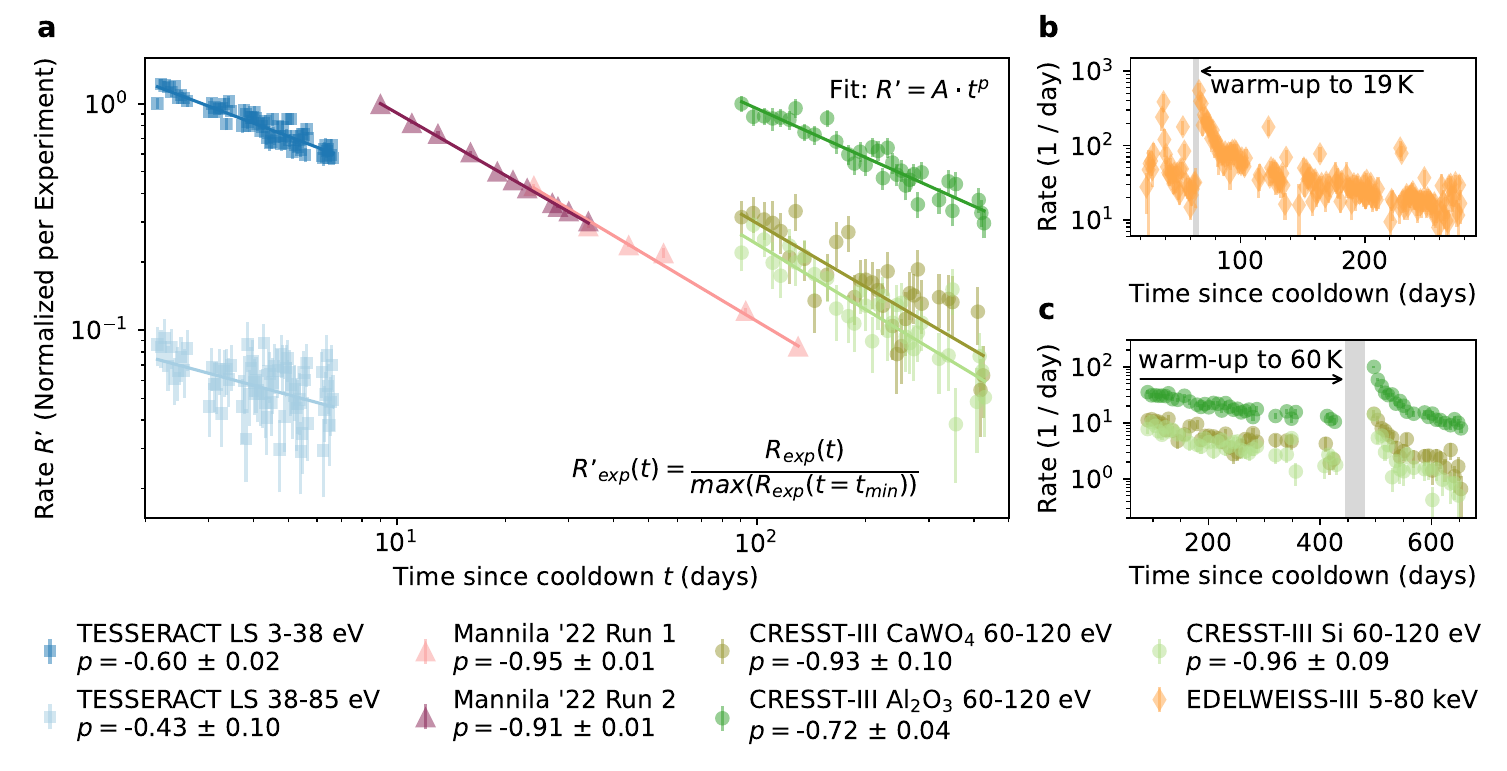}
    \caption{{\bf LEE rate evolution with time.} Low energy excess rates as a function of time since cooldown for various experiments: TESSERACT Low Stress (LS) configuration (Si, W-TES, surface)~\cite{anthonypetersen2022stressinducedsourcephonon}, Mannila (Al, quasiparticles, surface)~\cite{Mannila2022}, CRESST-III (CaWO$_4$/Si/Al$_2$O$_3$, W-TES, LNGS)~\cite{CRESST_LEE_2023}, EDELWEISS-III (Ge, NTD, LSM)~\cite{queguiner_edelweiss_thesis_2018}. \textbf{(a)} LEE rates in experiments with sufficient exposure decrease over time. The measured rates $R'$ are analyzed for each dataset separately and fit to a power law, shown as solid lines and indicated in the legend. The normalization is floated such that the rate at the earliest time point in the dataset with the highest rate for the corresponding experiment is set to $R' = 1$. 
    \textbf{(b)} Increase in the rate of heat-only events with energy ranging from 5 to 80\,keV, observed in an EDELWEISS-III detector following a warm-up to $\sim 19$\,K. \textbf{(c)}~Segment of a dedicated temperature study by CRESST, showing an enhancement in the LEE rate after a controlled warm-up to 60\,K.
    }
    \label{fig:time_decay}
\end{figure}

Early work by the EDELWEISS Collaboration~\cite{queguiner_edelweiss_thesis_2018} and more recent studies by CRESST~\cite{CRESST_LEE_2023} have demonstrated that the rate of LEE backgrounds in a given energy bin decreases over time. Subsequently, further observations of decaying LEE rates have been reported in various detectors~\cite{anthonypetersen2022stressinducedsourcephonon, phdthesis_Underwood_2022, CRESST:2TES2024}. This effect is illustrated in Fig.~\ref{fig:time_decay}a, where the LEE rates in several experiments are shown against time since cooldown. For CRESST, measurements were performed using different absorber materials, while for TESSERACT, the data are split into two energy ranges. In addition to these TES-based phonon detectors, we include data on the rate of Cooper pair-breaking events in a mesoscopic superconductor device~\cite{Mannila2022}. 
The data of the TES-based detectors are well described by exponentially decaying functions, as shown in the original studies~\cite{CRESST_LEE_2023, anthonypetersen2022stressinducedsourcephonon}. However, the resulting decay times vary significantly, ranging from several days for data taken immediately after the cooldown, to hundreds of days for those taken several months later. To account for this variability, we fit all experimental data with a power-law function $\propto t^p$. For the phonon detector data, both functions fit equally well, possibly due to the large spread of the data points. The data from quasiparticle tunneling can only be described well with a power-law model. Overall, the power-law function provides a consistent description across all considered experiments.
The resulting power-law exponent $p$ (see labels in Fig.~\ref{fig:time_decay}a) ranges from $-0.43$ to $-0.96$, with a weighted mean of $-0.9 \pm 0.1$.  
Whether time is defined from the start of the cooldown or from reaching base operating temperature substantially influences these fit results. 
Within the same cooldown, variations are observed in LEE rates across different energy ranges ({\it e.g.}, a faster decrease at lower energies in the TESSERACT detector) and in different absorber materials ({\it e.g.}, a slower decrease in CRESST Al$_2$O$_3$ vs. CaWO$_4$ and~Si). 

EDELWEISS observed an increase in HO events after a warm-up of the cryostat to $\sim 19$\,K~\cite{queguiner_edelweiss_thesis_2018}, as shown in Fig.~\ref{fig:time_decay}b. A series of warm-up tests were later performed by CRESST, partially shown in Fig.~\ref{fig:time_decay}c. Despite strongly varying energy ranges (above 5\,keV for EDELWEISS and $60-120$\,eV in CRESST), both collaborations observed that warm-ups to tens-of-K re-enhance the LEE rate, which begins to decrease again with time once detectors return to their operational temperatures. The rate decay following such warm-ups appears to be faster than the initial decay with time since cooldown. 

This behavior can be interpreted in different ways. One possibility is that warm-up procedures introduce transient effects that dissipate over shorter time scales. Alternatively, a faster decrease shortly after a detector reaches its operating temperature is expected naturally if the LEE rate follows a power-law decay over time, as discussed earlier. 

The observations of the time decay of these phonon low energybackgrounds are intrinsically incompatible with all hypotheses of common particle interactions due to their decay with time since the cooldown of the setup, leaving as possible origins solid-state physics effects that were previously unaccounted for. Such effects could include the spontaneous or stimulated rearrangement of non-equilibrium atomic or electronic configuration within the detector or surrounding materials to a lower energy state and subsequent transfer of energy to the detector phonon system. We elaborate on such ideas below, including stress, defect states, and excess QP population. 

\subsection{Stress} \label{subsubsec:stress} 

\begin{figure}[ht]
    \centering
    \includegraphics[width=1.\linewidth]{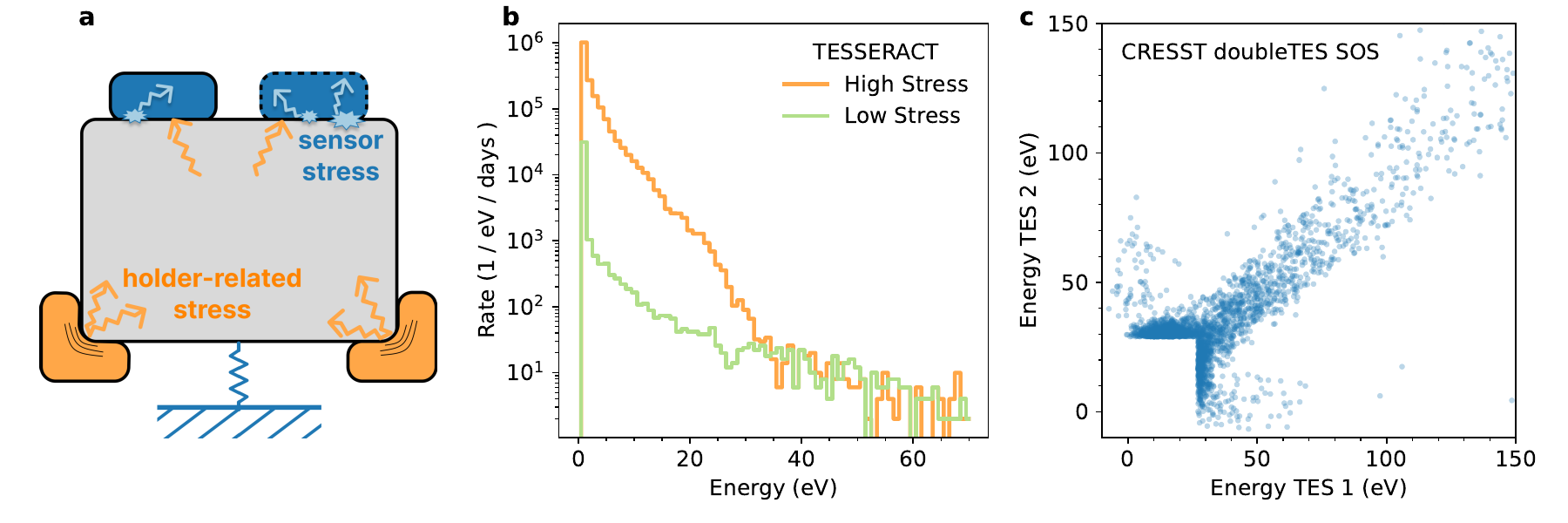}
    \caption{\textbf{Stress-related LEE origins.} \textbf{(a)} Schematic illustration of stress-related LEE origins, highlighting contributions from detector holders (orange) and phonon sensors (blue). 
    \textbf{(b)} Energy spectra measured with a Si TESSERACT detector~\cite{anthonypetersen2022stressinducedsourcephonon}, showing a higher LEE rate in the high-stress configuration compared to the low-stress configuration. \textbf{(c)} Scatter plot of energies measured by two TESs in a SOS double-TES CRESST detector operated at the surface~\cite{CRESST:2TES2024}, demonstrating the multi-component nature of the LEE: events detected by both sensors (shared) and events with energy deposition in only one sensor (singles).
    }
    \label{fig:stress}
\end{figure}

The cooldown of a cryogenic detector from room to mK temperature induces thermal contractions in all materials, and possibly shear stress on interfaces of components that contract differently. Since not all shear stress will immediately result in fractures or lattice reconfigurations, the remaining stress is a form of stored energy that can be released gradually over longer time scales, possibly inducing observable LEE events. Explaining discrete energy depositions of up to hundreds of eV via relaxation of shear stress through lattice reconfigurations requires the simultaneous rearrangement of many atoms, as a single-atom rearrangement is typically associated with smaller energy scales ({\it e.g.} $\sim 10$\,eV Frenkel defects in crystals). However, a comprehensive understanding and detailed description of this energy deposition mechanism remain to be developed. Systems with different material interfaces commonly found in phonon detectors (such as the crystal/holder or crystal/sensor film interfaces) are schematically illustrated in Fig.~\ref{fig:stress}a and discussed below.

\subsubsection{Detector holding stress}\label{subsubsec:holdingstress} 
Early work by the CRESST Collaboration demonstrated that crystal fractures caused by very tight clamping can spontaneously release energies ranging from keVs to several hundred keV~\cite{CRESST:Fractures2006}. More recently, TESSERACT showed that certain detector holding configurations are strongly correlated with LEE-type backgrounds~\cite{anthonypetersen2022stressinducedsourcephonon}. Specifically, the LEE rate was observed to be up to two orders of magnitude higher in a crystal operated under high stress, introduced by gluing it to a surface, compared to a crystal freely suspended on bond wires (see Fig.~\ref{fig:stress}b). 
More recent work by CRESST, in which a variety of sensor holding schemes were investigated, showed however that the observed low-energy background was insensitive to the exact holding method used~\cite{CRESST_LEE_2023}. 
These results suggest that stress caused by the mounting scheme may not be the dominant mechanism behind the low-energy background in CRESST detectors. However, the detector mounting process often lacks precise control over the level of induced stress.
Experimental architectures with active holding, such as the `i-sticks'~\cite{strauss_CRESSTprototype} previously used by CRESST, the instrumented Si inner veto developed by NUCLEUS~\cite{NUCLEUS1g:rothe2020}, and the segmented BULLKID monolithic detector array~\cite{BULLKID2022} are particularly well-suited for identifying holder-related stress events. Achieving sufficiently-low thresholds in all subsystems in such designs could provide more insights into the role of holding-induced stress in LEE formation.

\subsubsection{Sensor stress} \label{subsubsec:sensorstress}

The TESSERACT Collaboration suggested in Ref.~\cite{anthonypetersen2022stressinducedsourcephonon} that stress caused by differential thermal contraction of crystals and sensor metal films can be responsible for a part of the low energy phonon background. 
Subsequent work by  TESSERACT~\cite{TESSERACTTwoChannel}, CRESST~\cite{CRESST:2TES2024}, and NUCLEUS~\cite{2tes_nucleus_workshop24} focused on splitting the phonon readout into two channels, by attaching a second phonon sensor to the same crystal. 
These studies identified events depositing equal energy in both sensors (`shared' events) and those detected almost exclusively in one channel (`singles'). 
This distinction allows singles to be excluded from the final energy spectrum. 
Data from one such measurement with a CRESST detector~\cite{CRESST:2TES2024} are presented in Fig.~\ref{fig:stress}c.
These singles can be interpreted as events occurring in the metal films themselves~\cite{TESSERACTTwoChannel}, potentially stress-induced or caused by other mechanisms, such as bursts of microwave photons entering the sensor via the bias lines~\cite{TESSinglesGHzBursts}. 
In contrast, shared events are less likely to result from relaxation in the metal films, as such relaxation is expected to deposit at least some energy locally~\cite{Romani_Al_2024}. 
Instead, shared events are likely influenced by holder-related or intrinsic stress in the crystal, as well as stress at interfaces between the crystal and other instrumentation ({\it e.g.}, bond pads, thermal links, and electrodes). 

If stress relaxation in sensor components contributes to the observed LEEs, the specific materials responsible have yet to be identified.
Aluminum, widely used in phonon detectors, is a strong candidate~\cite{Romani_Al_2024}, but other common materials such as W, SiO$_2$, Au, and Nb may also play significant roles.  
However, comparing results across experiments remains challenging due to the need to isolate the sensor-related LEE component, compounded by variations in production methods, material sources, and the poor reproducibility often observed in thin-film fabrication.

\subsection{Defect states}

The lowest-energy state of a crystal lattice, the state of perfect symmetry, is in a realistic crystal never reached. Throughout the growth process, but also in any interaction of the lattice with its environment, lattice defects may accumulate. The relaxation of energy trapped in the crystal lattice in the form of defects has been hypothesized as a potential source of eV-scale energy depositions as phonon bursts~\cite{PhysRevD.105.063002}. 
Defects that were created through interface stress, as discussed in Sec.~\ref{subsubsec:stress} could act as such sights of stored energy. Also the impact of high-energy particles can create defects that create $\sim$eV-scale energy depositions in their relaxation, as studied in Ref.~\cite{DefectLEE}. 
Similarly to stress-relaxation induces backgrounds, also here an avalanche-like relaxation of many defects at once is required to explain the observed energies up to $\sim 100$\,eV. 
However, several questions remain unclear about that hypothesis, such as the long-term dynamics after the experiment's cooldown and the initial trigger for such an avalanche. 

Other types of defects, interpreted as trapped charges, in crystals are known to have defect structures whose configuration can resonate with microwave photons captured in quantum electronics and interact with electric fields and applied stress\cite{Lisenfeld2015, Bilmes2022}. Interestingly, their configuration is observed to drift over time and reset after thermal cycling  \cite{zanuz2024mitigatinglossessuperconductingqubits, reiss2024thermalcyclingevidence}.
Devices sensitive to macroscopic energy scales observe these defects as a bath and noise with $1/f$~frequency characteristic. 
The lifetimes of spin defects at low temperatures were observed to extend to months in CaWO$_4$ \cite{wang2024monthlonglifetimemicrowavespectralholes} and hours in diamond \cite{Astner2018}.
In addition, surfaces of many crystals are known to undergo chemical reactions over time, such as oxidation, or the adsorption of hydrogen at low temperatures. These processes could potentially release energy themselves, as is known {\it e.g.} from the adsorption of $^4$He atoms on Si \cite{PhysRevD.110.072006}, or change the likelihood of other reactions that deposit energy. We are not aware of studies of the evolution of surfaces of crystals on time scales of weeks in cryogenic low-pressure environments. A detailed study of the possible impact of all types of crystal defects on low energy backgrounds is beyond the scope of this review. 

\subsection{Excess quasiparticle population}
\label{sec:quasiparticles}

Many sources of athermal QPs are studied in modern literature, such as vibrations \cite{kono2023mechanicallyinducedcorrelatederrors}, environmental radioactivity \cite{Vepsaelaeinen2020}, and IR radiation \cite{10.1063/1.3638063}. 
These studies mostly focus on the impact that quasiparticles have on superconducting qubits, where they cause errors in quantum computations. 
While any interaction energy that exceeds the superconducting gap can create quasiparticles, some origins are more critical than others, due to high interaction rates or the creation of very high QP densities. 
The interaction of cosmic rays in the chip substrate~\cite{McEwen2022, Harrington:2024iqm} is of special interest as such backgrounds are difficult to shield and deposit enough energy to cause very high QP densities across the whole chip, and therefore correlated errors in multiple qubits that defeat error-correction algorithms. 
The absorption of IR photons through antenna modes of the circuit~\cite{PhysRevLett.132.017001} and QPs from stress in the substrate from the sample packaging~\cite{anthonypetersen2022stressinducedsourcephonon} have also been studied. The former one is especially critical due to its high interaction rates.
The dynamics of the QPs, once created, and their interaction with the qubit state, were described in Ref.~\cite{Martinis2021}.
Mitigation strategies for excess QPs have been studied, including normal-metal traps for athermal phonons \cite{Iaia2022} and QPs \cite{PhysRevB.94.104516}, gap engineering of a potential barrier at the Josephson junction \cite{mcewen2024resistinghighenergyimpactevents}, and shielding from cosmic rays through operation in underground laboratories~\cite{Cardani2021}.

It is expected that the origins of the LEEs observed in phonon detectors also create quasiparticles in quantum devices. The similar time-dependence of quasiparticle tunneling and data taken with TES, shown in Fig.~\ref{fig:time_decay}, points in this direction and is discussed further in Sec.~\ref{sec:discussion}.


\section{ORIGINS OF CHARGE EVENTS}

\begin{figure}[ht]
    \centering
    \includegraphics[width=1\textwidth]{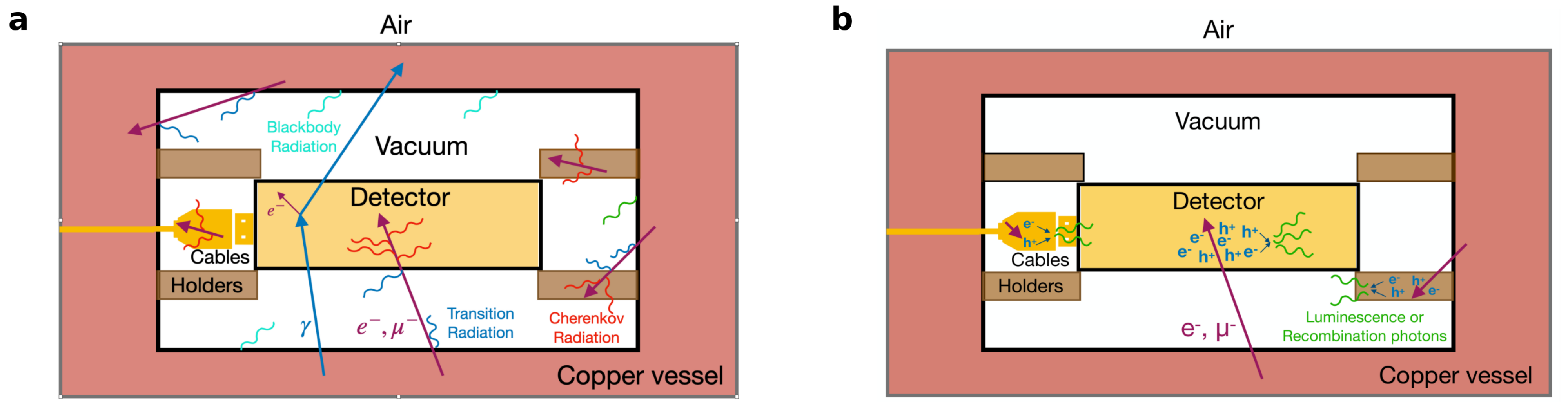}
    \caption{{\bf Charge backgrounds.} Production of charge backgrounds in detectors through ({\bf a}) primary radiation and ({\bf b}) secondary radiation. 
    High-energy charged particles (cosmic-ray muons or electrons) can either ({\it i}) pass through dielectric materials, such as the detector, the detector holders, or the cables, to produce Cherenkov radiation (red); ({\it ii}) cross surfaces to produce transition radiation (blue); or ({\it iii}) deposit energy in dielectric materials to create luminescence through, {\it e.g.}, the recombination of electron-hole pairs (green).  In addition, `warm' surfaces can produce blackbody radiation (cyan).  In all cases the resulting radiation can be absorbed in the detector to produce a low-energy signal. Compton scattering of high-energy photons (blue) can also produce low-energy electron recoils (purple). Figure closely adapted from Ref.~\cite{Du:2020ldo}.
    } 
    \label{fig:charge_contributions}
\end{figure}

\noindent Excess charge events can originate from primary or secondary radiation, as shown in Fig.~\ref{fig:charge_contributions}. These include Cherenkov radiation, luminescence, blackbody radiation, Compton scattering, transition radiation, thermal dark rates, and spurious charge.

\subsection{Radiation-induced backgrounds}
\label{ssec:radiation}

\subsubsection{Cherenkov} 
Charged particles (from cosmic rays or radioactive contaminants) passing through dielectric materials can emit low-energy Cherenkov radiation that is absorbed by the detector and can mimic a DM event. 
In Skipper-CCD detectors, charged-particle tracks can be masked, but Cherenkov photons can travel and be absorbed far from the tracks to produce a single-electron event~\cite{Du:2020ldo}, see Fig.~\ref{fig:radiation-induced}a.  
Two or more photons coincident in the same or neighboring pixels can mimic multi-electron events.  
Cherenkov radiation also explains the spatial correlation between charged tracks and single-electron events observed by SENSEI~\cite{SENSEI:2020dpa,Du:2023soy}.  
Cherenkov photons can also be produced in dielectric materials near the sensor and be absorbed by it.  This is a potential origin~\cite{Du:2020ldo} of the two- to six-electron events observed in SuperCDMS-HVeV data~\cite{SuperCDMS:2018mne}.  
Cherenkov radiation backgrounds can be suppressed by improving the shielding and reducing radioactive contaminants.

\subsubsection{Luminescence}
\begin{figure}[t]
    \centering
    \includegraphics[width=1\textwidth]{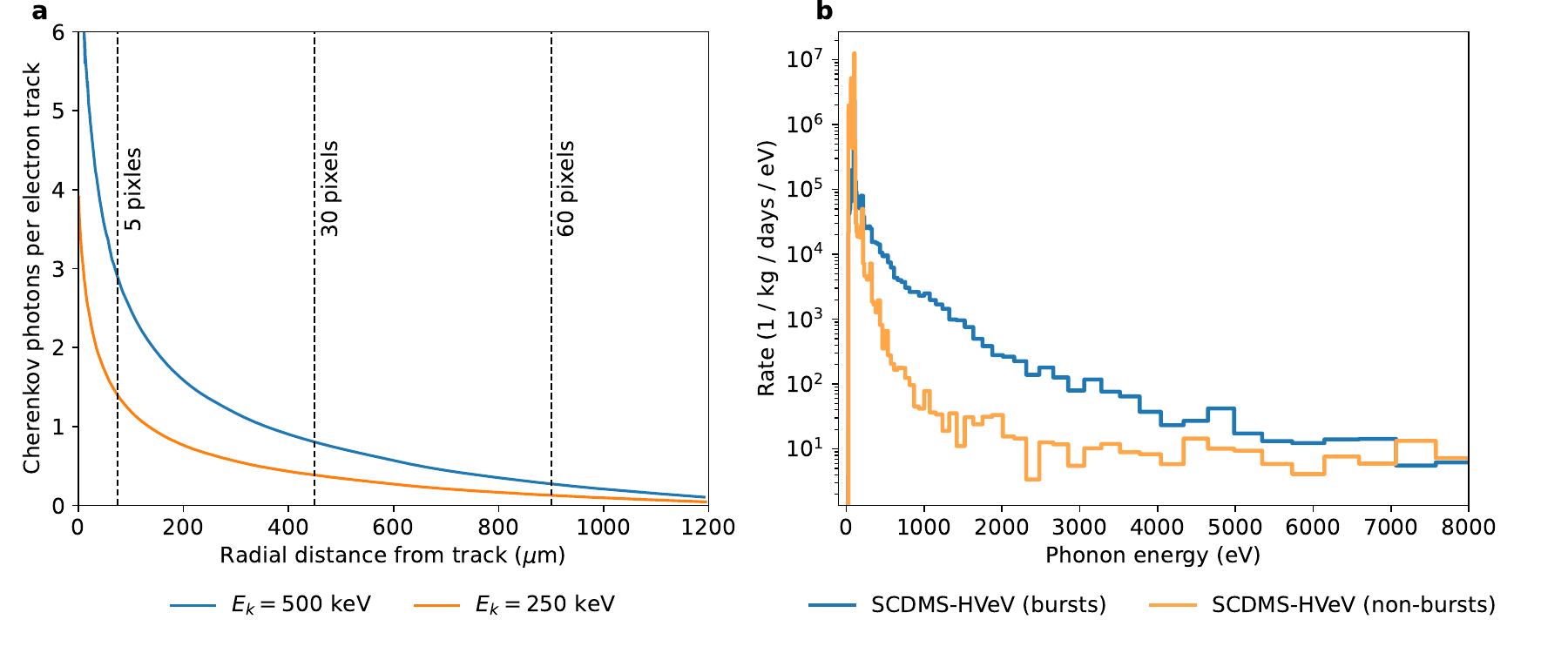}
    \caption{
    {\bf Charge-event background sources.} \textbf{(a)}~Number of Cherenkov photons emitted by one electron track passing through Si with energy  500\,keV (blue) and 250\,keV (orange) versus the radial distance from the track \cite{Du:2020ldo}. Vertical lines show the distance in pixels in Skipper-CCDs. 
    {\bf (b)}~Total phonon energy spectrum of burst events (blue) and non-burst events (orange) observed with SuperCDMS-HVeV operated at 100\,V \cite{SuperCDMS:2022zmd}. The reconstructed energy of the burst events is close to the energy of the primary pulse.
    }
    \label{fig:radiation-induced}
\end{figure}
Particles that traverse materials deposit energy in them, 
which may excite \eh pairs creating photons after recombination, a process commonly termed luminescence.
This process can be prompt or delayed.
As an example of delayed luminescence, SiO$_2$, the primary component of PCB as used in the original SuperCDMS-HVeV design (see Sec.~\ref{sssec:hvev}), can emit luminescence photons of 4.4, 1.9, and 2.7\,eV with a decay time of 1.5\,$\upmu$s, 20\,$\upmu$s, and 7\,ms, respectively \cite{SuperCDMS:2022zmd}.
These energies and timescales can account for the excess of SuperCDMS-HVeV events featuring closely packed pulses in time, referred to as `bursts' (see Fig.~\ref{fig:radiation-induced}b).
A burst consists of a distribution of primary events, forming a single high-energy primary pulse, and of secondary events quantized in one to three \eh pairs adding to the LEE.

\subsubsection{Infrared/Blackbody}
The energy thresholds of phonon and charge detectors are often well-within the regime of sensitivity to single optical photons and below, such that thermal radiation and other sources of IR photons can become a source of low-energy backgrounds. For detectors with higher thresholds, radiation of sub-eV photons is not resolvable on an event-by-event basis, but can still be visible as shot noise and deteriorate the energy resolution given a sufficient power flux.
Blackbody radiation has recently been identified by SENSEI as an important contributor to the single-electron event rate in data taken at SNOLAB and MINOS~\cite{SENSEI:2024yyt}.  Using a new design for the trays that hold the Skipper-CCD sensors with fewer light leaks, a reduced single-electron event rate was found. The hypothesis is that the warm surfaces inside the vessel that houses the trays emit blackbody radiation that can enter via light leaks and be recorded as a single-electron event.
Additionally, the CCD amplifier can emit IR light during readout, which can be absorbed elsewhere in the CCD~\cite{SENSEI:2019ibb}.  Amplifier glow can be mitigated by optimizing the readout voltages and the amplifier design, and turning off the amplifier voltages during exposure. 
Backgrounds compatible with room-temperature photons were furthermore observed with an optical TES in a sub-Kelvin cryogenic setup in Ref.~\cite{PhysRevApplied.22.024051}. Thermal IR radiation is also among the sources of high excess QP rates observed in superconducting qubits, see Sec.~\ref{sec:quasiparticles}.

\subsubsection{Compton scattering} Photons that undergo Compton scattering off electrons will produce electron recoils over a broad range in energies, including at the energies relevant for DM interactions~\cite{Botti:2022lkm,DAMIC-M:2022xtp,Essig:2023wrl}. Careful control over radioactive contaminants and other environmental radiation is crucial for reducing this background to negligible rates. We note that Compton scattering does not explain the LEE, as its shape is different, but it can produce low-energy events.

\subsubsection{Transition radiation}
Charged particles that traverse an interface that separates two different materials can produce transition radiation.  This can occur when traversing two dielectric media or even a copper-vacuum interface.  The number of photons emitted by a transition is far smaller than the number of Cherenkov photons produced by a charged particle traversing a dielectric material, but (unlike for Cherenkov radiation) there is no energy threshold for the charged particle. This radiation can be absorbed by a DM sensor and mimic a DM event~\cite{Du:2020ldo,Robinson:2020zec}.

\subsection{Fundamental backgrounds}
\subsubsection{Dark rates (thermal)}
True dark rates are generated from thermal fluctuations even if the detector is in complete darkness and perfectly isolated from its environment.  
In a CCD, there are various possible sources, including surface dark current, depletion dark current, bulk dark current, and diffusion dark current~\cite{janesick2001scientific}.  
High-quality Si can reduce the bulk dark current dramatically, while surface dark current can be reduced through careful choice of CCD operational parameters.  
Nevertheless, some dark current will remain at finite temperatures and is exponentially suppressed for lower temperatures.  
The general dark current formula for a Si CCD with pixel area $A$, dark current `figure of merit' $D_{\rm FM}$, band gap $E_{\rm gap}$, operating at a temperature $T$ (in Kelvin) is given by~\cite{janesick2001scientific} 
\begin{equation}\label{eq:thermal}
    D_R \simeq 5\times 10^{13} \frac{e^-}{{\rm pixel~day}}\, \frac{A}{(15~\upmu{\rm m})^2} \, \frac{D_{\rm FM}}{0.1~{\rm nA}}\, T^{3/2} \, e^{-\frac{E_{\rm gap}}{2 k_B T}}\,,
\end{equation}
where $k_B$ is the Boltzmann constant. 
For illustration purposes, we use $D_{\rm FM}=1$~pA as this is roughly the value needed to match the upper limit set by the bulk dark rate in Fig.~10 of Ref.~\cite{SENSEI:2020dpa}, although it is most likely an overestimate of the true thermal dark rate (this is the dashed line shown in Fig.~1). 
In any case, we see that the dark current is very small for typical CCD operating temperatures. 
Thermal dark rate is negligible for a cryogenic detector operated below a few Kelvin. 

\subsubsection{Spurious charge}%
Spurious charge in a CCD is generated when clocking the pixels, {\it i.e.}, when 
transferring charges between the pixels~\cite{janesick2001scientific}.  It is spatially uniform and independent of the exposure, and the accumulated spurious charge depends on the number of times a pixel is clocked. Its average contribution can be subtracted from the exposure-dependent single-electron contribution~\cite{SENSEI:2021hcn}, and it can be reduced by optimizing the clock voltages.

\section{DISCUSSION AND OUTLOOK} \label{sec:discussion}

\begin{figure}[ht]
    \centering
    \includegraphics[width=\linewidth]{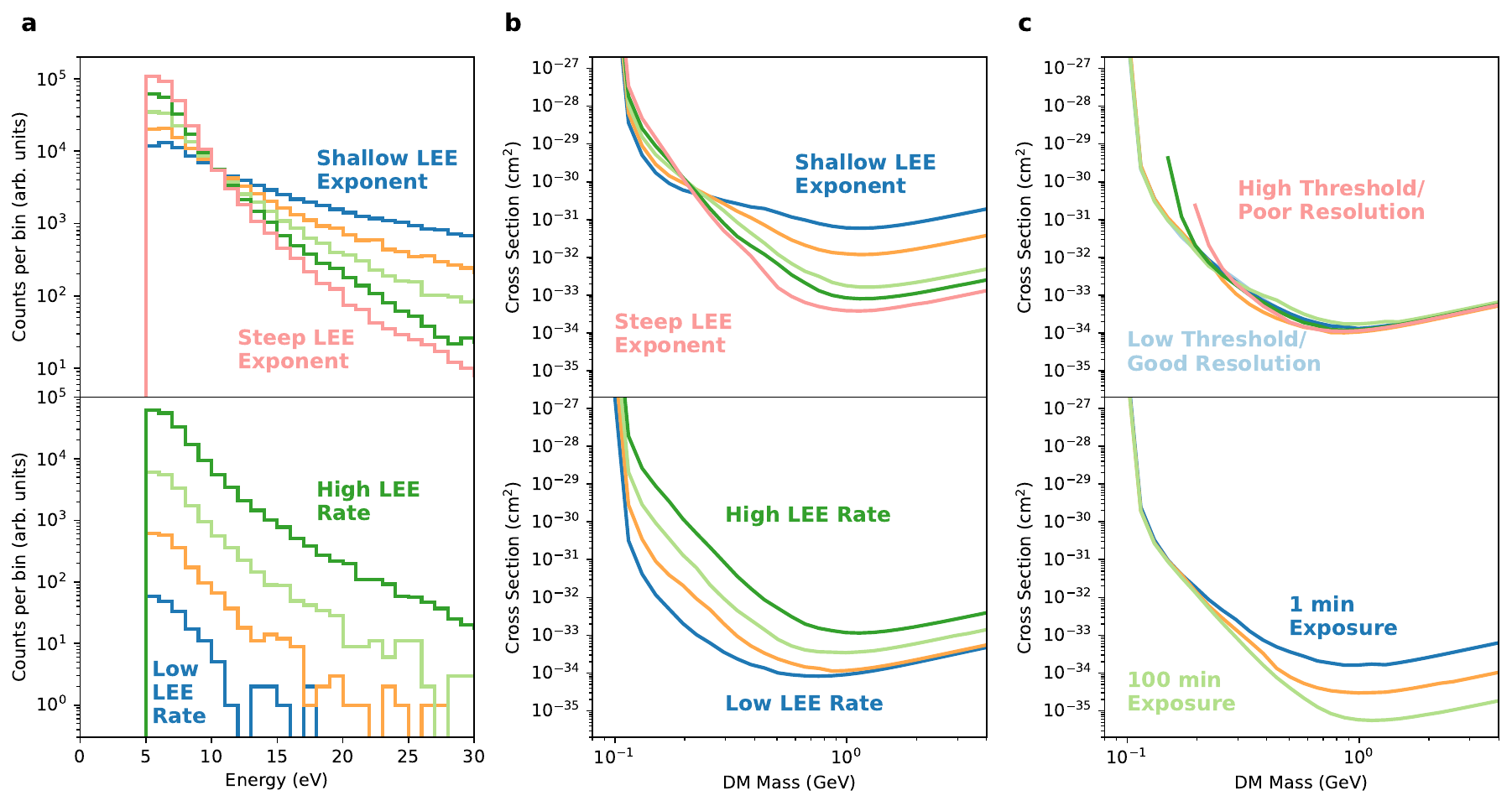}
    \caption{ {\bf Simulations of LEEs impact on DM limits.} Spin-independent DM-nucleon projected limits for a 1\,g Si detector, assuming a single power law background $ r_0 (\frac{E} {\mathrm{1 eV}})^{-\kappa}$. Unless otherwise stated, projected limits are for a detector with a resolution $\sigma = 1$\,eV, a threshold of 5\,eV, and an exposure of 60\,s. {\bf (a)} {\it Top \& Bottom:} Simulated event spectra, varying the power law and rate terms together ($\kappa = [2, 3, 4, 5, 6]$ , $r_0 = [10^2, 10^3, 10^4, 10^5, 10^6]$\,Hz/eV at 1\,eV) and rate ($\kappa = 5$, $r_0 = [10^2, 10^3, 10^4, 10^5]$\,Hz/eV at 1\,eV) respectively. {\bf (b)} {\it Top \& Bottom: } Simulated projected limits given the backgrounds in  {\bf (a)}, with a 60\,s exposure of a 1\,g silicon detector. Coloring is consistent in {\bf (a)} and {\bf (b)}, {\it e.g.} in top plots orange lines correspond to $\kappa = 3, r_0 = 10^3$\,Hz/eV events. {\bf (c)} {\it Top:} Simulated projected limits in a detector with a varying resolution $\sigma = [0.2, 0.3, 0.5, 1, 2, 3]$\,eV, a threshold of $5\sigma$, and a background of $\kappa = 5, r_0 = 10^4$\,Hz/eV. {\bf (c)} {\it Bottom:} Simulated limits with exposures of [1, 10, 100]\,min.}
    \label{fig:excesses_sim}
\end{figure}

\noindent Understanding and mitigating LEEs is an extremely important task for further sensitivity improvements of low-threshold DM and \cevns experiments, as illustrated in Fig.~\ref{fig:excesses_sim} for the case of the DM nuclear recoil exclusion power. 
The main takeaways are: ({\it i})~The shape of the LEE matters. 
A steep LEE will be more similar to the spectrum of a light DM particle and thus have a higher impact at low DM masses and vice versa (Fig.~\ref{fig:excesses_sim}a\&b {\it top}).  
({\it ii})~The rate of the LEE is crucial. 
This is demonstrated by Fig.~\ref{fig:excesses_sim}b {\it bottom} showing that an elevated LEE negatively impacts the sensitivity over the whole DM mass range. ({\it iii})~ For background-limited/LEE-dominated experiments, a lower threshold (Fig.~\ref{fig:excesses_sim}c {\it top}) or higher exposure (Fig.~\ref{fig:excesses_sim}c {\it bottom}) might not improve the sensitivity. 
As expected, a lower threshold extends the sensitivity to lower DM masses, but the improvements for higher masses are very modest.  
Similarly, increasing the exposure will not help for low DM masses where the sensitivity is limited by the LEE. 

Both spin-independent DM-nucleus scattering and \cevns have cross-sections that scale coherently, assuming an equal number of protons and neutrons in the target nuclei. 
Since the energy spectrum produced by \cevns from reactor neutrinos closely resembles that expected from a 2.7\,GeV/c$^2$ DM particle~\cite{billard2021}, the neutrino flux corresponds to the equivalent DM interaction cross-section and the LEE's impact on \cevns detectors' sensitivity can be inferred from Fig.~\ref{fig:excesses_sim}. 
Similar statements apply to sensitivities for DM-electron scattering. 
For example, 1\,MeV DM particles produce only single-electron events (as the transferred energy is below the two-electron threshold), and since no experiment has yet succeeded in subtracting these backgrounds, DM limits scale linearly with reductions in single-electron rates.
Even for DM heavier than a few MeV, the signal predominantly consists of only a few electrons, unlike DM-nucleus scattering. 
Reducing one- and few-electron backgrounds will directly enhance sensitivity to sub-GeV DM.

One of the key findings of the concerted effort to understand and characterize the LEE is that the dominant sources appear to differ between phonon- and charge-sensitive detectors.
However, the nature of these underlying sources remains not fully understood.
For phonon-sensitive detectors, compelling similarities are observed across very different experiments, including the energy scales at which the excesses occur (with the notable exception of EDELWEISS), as well as the time-dependence and re-enhancement of LEE rates after warm-up. 
Detectors with double-TES readout, operated by CRESST, NUCLEUS, and TESSERACT, exhibit strikingly similar distributions of double- and single-TES events~\cite{CRESST:2TES2024, 2tes_nucleus_workshop24, TESSERACTTwoChannel}, albeit at different energy scales. 
These observations rule out a particle (signal and/or background) origin for the LEE.

Moving forward, the path toward DM and \cevns~science is best pursued through a program of diverse small-scale R\&D, combining focused efforts to test individual LEE models with broader initiatives to explore LEE in new regimes or with new tools. 
A set of LEE models exists with fairly well-defined predictions. For example, in the radiation-induced crystal defect relaxation model~\cite{DefectLEE}, the rate of LEE in the energy range of interest should significantly increase after or while the detector is exposed to large amounts of defect-creating radiation. In the aluminum relaxation model~\cite{Romani_Al_2024}, the LEE rate should be lower for detectors with less aluminum compared to those with more.
Certainly, the observed decrease in the LEE rates over time at operating temperature is one of the best hints as to the nature of the LEE. A detailed investigation of LEE rates across different time scales, along with further studies on temperature dependence and impact of the cooldown speed, may provide more insights on the underlying processes. From the currently available observations, a set of ideas focused on the relaxation of stress induced on the crystal or at its interfaces with other detector components has emerged.

To address the stress on the target crystal caused by holding structures, two main strategies are being employed.
The first relies on instrumenting the holding structures and rejecting holder-related events through a dedicated veto cut.  NUCLEUS is developing a Si holding structure equipped with a TES~\cite{Nucleus2023}.~Similarly,~CRESST  has adopted an active holder approach by gluing the target crystal to another instrumented crystal in its mini-beaker module design~\cite{CRESST:DetDev2023}. The second approach focuses on reducing external pressure on the target crystal, as implemented by TESSERACT with crystals supported solely by bond wires~\cite{anthonypetersen2022stressinducedsourcephonon} and by CRESST using a gravity-assisted holding scheme~\cite{CRESST:DetDev2023}.

Detectors with double-TES readout revealed a fraction of low energy background events localized in a single phonon sensor, which can be identified and removed from the final energy spectra~\cite{TESSERACTTwoChannel,CRESST:2TES2024,2tes_nucleus_workshop24}. Combining instrumented holders with the double-readout approach, as planned by the NUCLEUS experiment, might enable the identification of both sensor-related and holder-related LEE events.
Stress-related LEEs are also expected to~be suppressed for superfluid He targets read out by solid-state phonon-based microcalorimeter arrays, since the target He cannot experience any stress. 
LEE events originating from stress in the microcalorimeters will be reduced by exploiting the mostly-different hit patterns between particle events and stress-induced events.
This approach is being taken by the HeRALD~\cite{PhysRevD.110.072006} and DELight~\cite{delight:2022} experiments.

Two new NTD-based detector designs have evolved from EDELWEISS, called CRYOSEL (used by EDELWEISS~\cite{Lattaud:2023wyc}) and CryoCube (used by RICOCHET \cite{Ricochet2023}). 
The CRYOSEL-CRYO50 prototype is a Ge crystal equipped with two sensors, an NTD thermal phonon sensor and a NbSi film sensitive to athermal phonons. The latter is designed to measure charge through the detection of NTL phonons and, thus, enables the identification of bulk ionizing events and consequently the rejection of HO background events, which are a major component of the LEE in EDELWEISS (see Sec.~\ref{sssec:edelweiss}).
Initial tests with IR laser pulses on CRYO50 demonstrate the potential of this sensor technology to achieve the crucial single \eh pair sensitivity~\cite{Lattaud:2023wyc}.

Another promising approach to shed light on the origin of sources for the LEE is provided by pulse-shaped discrimination, as predicted, for example, by MAGNETO-DM \cite{magneto-dm-2024}.
MAGNETO-DM uses a diamond target with a Au phonon absorber instrumented with a magnetic microcalorimeter (MMC) \cite{Ens00} to collect athermal phonons after an energy deposition.
They exploit the fast response times achievable with MMCs 
to distinguish between energy depositions in the crystal, the Au phonon absorber, and the MMC sensor, in addition to discriminating between electronic and nuclear recoil events \cite{magneto-dm-2024}. 
This will enable a characterization of events contributing to the LEE, observed in recent MAGNETO-DM data, and is expected to provide new insights.

These and other upcoming measurements may yield new insights, particularly clarifying whether each LEE background is driven by a single mechanism or by a small set of related effects.
However, the shared characteristics of the observed LEEs may well arise from a semi-universal property of condensed matter systems at low temperatures, with variations specific to each experiment.
The availability of accurate models of the low-energy backgrounds can regain the ability to search for new physics, especially when observations can be compared across data sets from multiple detectors, as was shown {\it e.g.}~in Ref.~\cite{Wagner2023}. 
In the event that these backgrounds are intrinsic to cryogenic solid-state systems, a path forward may rely on the development of low-threshold sensors with directional discrimination between isotropic LEE backgrounds and a DM or \cevns~signal~\cite{Blanco:2021hlm}. 

Parallel observations in quantum circuits made of similar materials and operating in similar environments (see {\it e.g.} Refs.~\cite{PhysRevB.110.024519,Mannila2022}) have revealed excess event rates with similar characteristics.
The bursts of excess QPs observed in an Al island on Si \cite{Mannila2022} resemble the phonon excess background seen with TES detectors with Al absorbers on Si \cite{PhysRevD.107.122003}. Both show similar time decay rates (Fig.~\ref{fig:time_decay}) for recoils above 60\,eV. Extrapolating the TES observations to lower energies, using an energy spectrum $\propto$(energy)$^{-2.2}$, yields $\sim$\,Hz recoil rates in the $100–1000$ $\times$ Al gap ($\sim 0.18$\,meV). Assuming a $\sim 1$\% phonon-to-QP collection efficiency (in the range of efficiencies cited in Refs.~\cite{Proebst1995, 10.1063/5.0032372, PhysRevLett.133.060602}), these rates align with QP bursts in the Al island. This suggests a possible shared origin, though alternative explanations and other contributions, such as from thermal IR photons, cannot be ruled out. 
The origins and mitigation of excess quasiparticles are the focus of ongoing research given their impact on superconducting quantum computers.

For charge-sensitive detectors, Fig.~\ref{fig:excess_rates}b  shows how  the single-electron (and two-electron) background rates have steadily been reduced. This has been achieved through a combination of characterizing their origins, improving the operation of the detectors, and reducing the environmental backgrounds. Indeed, several origins of low-energy charge events have now been linked to (high-energy) environmental backgrounds, including Cherenkov radiation, luminescence, Compton scattering, and transition radiation, and can be reduced through careful detector shielding and radiological control.
Blackbody radiation can be reduced by removing light leaks in the sensor housing to avoid light from warm surfaces reaching the sensor. Dark rates from thermal excitations can be reduced by cooling the detector, and the expectation is that these can be reduced to negligible levels even for the CCD detectors, which need to be operated above 100~K. 
Infrared light from the CCD amplifier can and must be controlled.  Spurious charge generated during CCD readout produces an exposure-independent single-electron rate, which can be separated statistically from an exposure-dependent rate generated by DM; however, additional work is needed to reduce it further to maximize the detector's sensitivity to DM.  

SENSEI's next steps include reducing the spurious charge and the blackbody radiation. For the latter, an updated light-tight design of the CCD copper trays will be employed. SENSEI's goal of accumulating an exposure of $\mathcal{O}$(100~gram-year) requires some control over radiogenic backgrounds, well within what has been demonstrated. 
DAMIC-M aims to obtain an exposure of 1~kg-year, which requires tighter control over radiogenic backgrounds than in SENSEI to avoid LEE rates in the DM signal range. 
Both experiments will generate sufficient exposure to independently probe the origins of the DAMIC at SNOLAB excess rates, which appear distinct among LEEs discussed here. 
Moving forward, OSCURA \cite{2022arXiv220210518A} will merge the global CCD effort for an exposure of $\sim$30~kg-year, placing stringent requirements on avoiding radioactive contaminants and environmental exposure. As part of this, the CCDs will be immersed in liquid nitrogen, which will help reduce blackbody radiation.

SuperCDMS-HVeV has recently operated multiple HVeV detectors at the CUTE underground facility at SNOLAB with a detector design and placement dedicated to the characterization of the LEE.
Six HVeV detectors are stacked in a light-tight tower, with  four of them in the same optical cavity.
Two detectors have a new design which includes ultra-thin insulating SiO$_2$ layers to inhibit charges tunneling through the interfaces between deposited metals and Si to reduce the single \eh pair rate.
The aluminum coverage also varies between some of the detectors in order to probe the impact of this quantity on the LEE.
The results will inform the next steps in the SuperCDMS-HVeV program.

Years after their first observation, LEEs continue to limit the scientific reach of experiments using low-threshold detectors. However, notable progress has been made in understanding them, alongside continued, albeit slowed, scientific advancements. Uncovering the origins of LEEs will likely require collaboration across the material science, solid-state physics, and quantum science communities, fostering an open-minded approach to low-energy phenomena. In our view, the LEE community should remain optimistic that backgrounds at low energies will eventually be understood and mitigated through ongoing projects, restoring sensitivity to interactions at the eV-scale and beyond. Furthermore, a deeper understanding of our detectors and the fundamental processes involved—an inevitable outcome of these efforts—may inspire new technological advancements.


\section*{DISCLOSURE STATEMENT}
The authors are not aware of any affiliations, memberships, funding, or financial holdings that
might be perceived as affecting the objectivity of this review. 

\section*{ACKNOWLEDGMENTS}
The authors are grateful for discussions with all contributors to the EXCESS Workshop Series and to each of the collaborations whose data and findings we summarize in this review. 
This manuscript has been co-authored by Fermi Research Alliance, LLC under Contract No.\,DE-AC02-07CH11359 with the U.S. Department of Energy (DOE), Office of Science, Office of High Energy Physics.
DB is supported by DOE, Office of Science, National Quantum Information Science Research Centers, Quantum Science Center.
RE is supported from DOE Grant DE-SC0025309, Simons Investigator in Physics Awards~623940 and MPS-SIP-00010469, Heising-Simons Foundation Grant No.~79921, and Binational Science Foundation Grant No.~2020220. 
YH is supported by the Israel Science Foundation (grant No. 1818/22), by the Binational Science Foundation (grant  No.\,2022287) and by an ERC STG grant (``Light-Dark'', grant No.\,101040019).
MK is supported by the Deutsche Forschungsgemeinschaft (German Research Foundation, DFG) under Germany’s Excellence Strategy – EXC-2094 – 390783311, through the Collaborative Research Center SFB1258 "Neutrinos and Dark Matter in Astro- and Particle Physics", by BMBF Grant No. 05A23WO4, and by the European Commission through the ERC-StG2018-804228 "NU-CLEUS".
BvK is supported by the DFG under the Emmy Noether Grant No.\,420484612. 
RKR is supported in part by DOE Grants DE-SC0019319, DE-SC0022354, DE-SC0025523 and DOE Quantum Information Science Enabled Discovery (QuantISED) for High Energy Physics (KA2401032). 
This project has received funding from the European Research Council (ERC) under the European Union’s (EU) Horizon Europe research and innovation programme (grant No.\,101040019).  Views and opinions expressed are however those of the author(s) only and do not necessarily reflect those of the EU. The EU cannot be held responsible for them. 

\bibliographystyle{ar-style5}
\bibliography{bibliography}

\end{document}